\documentclass[a4paper,11pt]{article}
\usepackage{pos}
\title{Overview and First Results of EUSO-SPB2}
 \ShortTitle{EUSO-SPB2 Overview}

\author*[a]{Johannes Eser}
\author[a]{Angela V. Olinto}
\author[b]{Lawrence Wiencke}

\affiliation[a]{The University of Chicago, Department of Astronomy \& Astrophysics,\\
  5640 South Ellis Avenue, Chicago, IL, USA}

\affiliation[b]{Colorado School of Mines, Department of Physics,\\
1523 Illinois Street, Golden, CO, USA}

\onbehalf{for the JEM-EUSO collaboration} 

\emailAdd{jeser@uchicago.edu}
\emailAdd{aolinto@uchicago.edu}
\emailAdd{lwiencke@mines.edu}

\abstract{
Observing ultra-high energy cosmic rays (UHECR) and very high energy (VHE) neutrinos from space
is a promising way to measure their extremely low fluxes by significantly increasing the observed
volume. The Extreme Universe Space Observatory on a Super Pressure Balloon 2 (EUSO-SPB2), the next, most advanced pathfinder for such a mission, was launched May 13th 2023 from Wanaka New Zealand. The pioneering EUSO-SPB2 payload flew a Fluorescence Telescope (FT) with a PMT camera pointed in nadir to record fluorescence light from cosmic ray extensive air shower (EAS) with energies above 1 EeV, and a Cherenkov telescope (CT) with a silicon photomultiplier focal surface for observing Cherenkov emission of cosmic ray EAS with energies above 1 PeV with an above-the-limb geometry and of PeV-scale EAS initiated by neutrino-sourced tau decay. As the CT is a novel instrument, optical background measurements for space neutrino observation are an important goal of the mission. Any data collected during the mission will influence and improve the development of a space-based multi-messenger observatory such as the Probe of Extreme Multi-Messenger Astrophysics (POEMMA). We present an overview of the EUSO-SPB2 mission and its science goals and summarize results as available, from the 2023 flight.

}

\ConferenceLogo{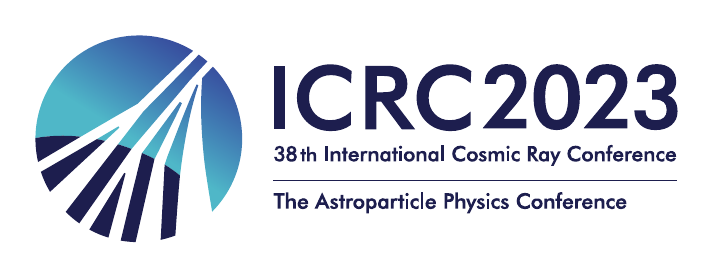}

\FullConference{%
38th International Cosmic Ray Conference (ICRC2023)\\
  26 July - 3 August, 2023\\
  Nagoya, Japan}

\begin{document}
\maketitle
\section{Introduction}
\label{sec:intro}

The Extreme Universe Space Observatory on a Super Pressure Balloon 2 (EUSO-SPB2) was a pathfinder for the Probe of Multi-Messenger Astrophysics (POEMMA) \cite{POEMMA-JCAP}. POEMMA is a proposed dual satellite mission for the detection of ultra-high energy cosmic rays (UHECRs, E>1~EeV) and very-high energy neutrinos (VHENs, E>1~PeV). These particles, the most energetic ones known to exist, offer a unique opportunity to study the most extreme astrophysical events in the universe and to understand the cosmological evolution of sources and production and acceleration mechanisms of these particles. A detailed discussion of the current status and the future of UHECR, including the role of multi-messenger Astrophysics, is given in \cite{COLEMAN2023102794}. The flux of cosmic rays and neutrinos decreases rapidly with energy. UHECRs and VHENs both require a large instrumental volume to accumulate sufficient statistics to discover sources and study them. POEMMA addresses this challenge by viewing Earth and its atmosphere from the vantage point of space with a wide field view, dramatically increasing the geometric acceptance relative to ground-based experiments. POEMMA will observe neutrinos by measuring the Cherenkov light emitted by extensive air showers (EAS). These upward-moving EAS are produced by the interaction or decay of a charged lepton in the atmosphere after a neutrino propagates through the Earth and interacts near the surface. UHECRs are detected via the fluorescence light emitted by the EAS, which is produced through the interaction of the primary cosmic ray with the atmosphere.

\begin{figure}[!ht]
\centering
\includegraphics[width=1.\textwidth]{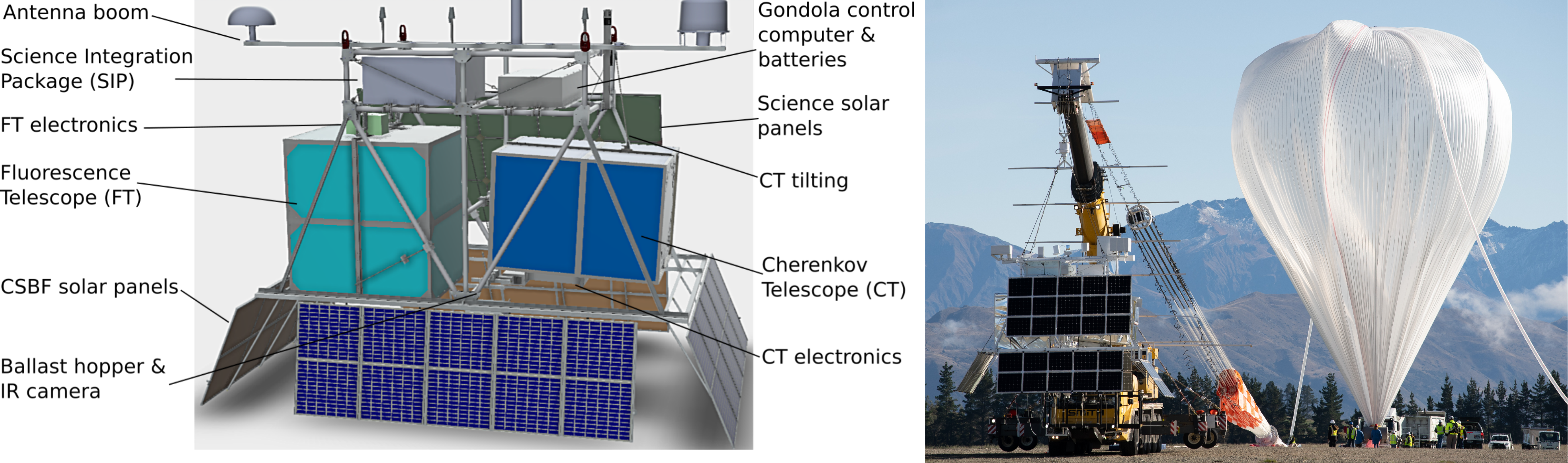}
\caption{A labeled sketch of EUSO-SPB2 (left) and a picture of the fully assembled instrument before launch (right). The sketch shows the two telescopes (FT and CT) and the IR cloud camera, as well as the long-duration balloon flight gear, such as solar panels and telemetry antennas. The picture shows the instrument hanging from the launch vehicle. (Credit: B. Rodman)}
\label{fig:payload}
\end{figure}

While each POEMMA telescope features a hybrid focal surface design to detect both signals, the functions of the two telescopes of the EUSO-SPB2 pathfinder balloon payload were separated between a Fluorescence telescope (FT) and a Cherenkov telescope (CT) (fig. \ref{fig:payload}). The main goal for the FT was to observe UHECRs from above via the fluorescence technique for the first time. The CT goals can be divided based on its pointing direction; when pointed above the limb, it aimed to measure cosmic rays with the direct Cherenkov technique for the first time, and when pointed below the limb, it could measure optical background signals for neutrino searches and  search for neutrinos signatures from astrophysical targets of opportunity. The increase technology readiness levels can be applied to POEMMA or any other future space mission. 

EUSO-SPB2 launched on May 13th, 2023, from Wanaka, NZ, as a mission of opportunity on a NASA super pressure balloon that terminated prematurely after only 1 day, 12 hours 53 minutes at float in the Pacific Ocean due to a leak in the balloon. This extremely short flight, compared, for example, to the 39 day flight of the other mission launched in this campaign was unfortunate. However, the data collected demonstrates that all the instruments turned on successfully at float altitude and performed well during the highly-curtailed flight opportunity. Some of the scientific and technical goals were realized.
\section{The Fluorescence Telescope (FT)}
\label{sec:FT}

\textbf{The instrument}, a 1m diameter modified Schmidt telescope, was designed and built to measure the fluorescence light produced by the particles of an EAS induced by a UHECR. Its camera has a modular design based on MultiAnode Photo Multiplier Tubes (MAPMTs: R11265-M64-203 from Hamamatsu) operated by a custom application-specific integrated circuit \cite{SPACIROC3} that allows for single photo electron counting with a double pulse resolution of 6~ns and an integration time of 1$\mu$s. Four of these tubes form an Elementary cell (EC), which shares one Cockcroft-Walton board to apply the required HV of 1050V. Nine ECs make up a Photo Detection Module (PDM). Each PDM is covered with a BG3 filter \footnote{\url{https://www.schott.com/shop/advanced-optics/en/Matt-Filter-Plates/BG3/c/glass-BG3}} to limit its sensitivity to the UV region (290-430~nm) and reduce the background of photon counts outside this optimal frequency band. 
The trigger \cite{FILIPPATOS20222794} operates individually on each PDM. The 6912-pixel camera consists of 3 such PDMs. Its resulting field of view (FoV) is 36 by 12 degrees with a fixed nadir pointing direction. A central data processor \cite{SPB2_DP} connects the 3 PDMs and enables a synchronized readout if one PDM triggers. In addition to combining the photo count data of all PDMs, it adds GPS information to each event and prioritizes the events for download using a neural network trained on simulation before the flight (see \cite{GeorgeICRC_ML}).

\textbf{The expected performance} was studied via Monte Carlo simulations of an isotropic flux of UHECRs (for details, see \cite{gfil-ICRC}) using the JEM-EUSO OffLine framework \cite{Offline}, which provides an end-to-end simulation of the response including atmospheric effects, background signals based on previous flights and the trigger. These simulations showed a peak energy sensitivity of around 2.5~EeV. Across all energies, the expected event rate was one event approximately every five to eight hours of clear observation.

\begin{figure}[!ht]
\centering
\includegraphics[width=.7\textwidth]{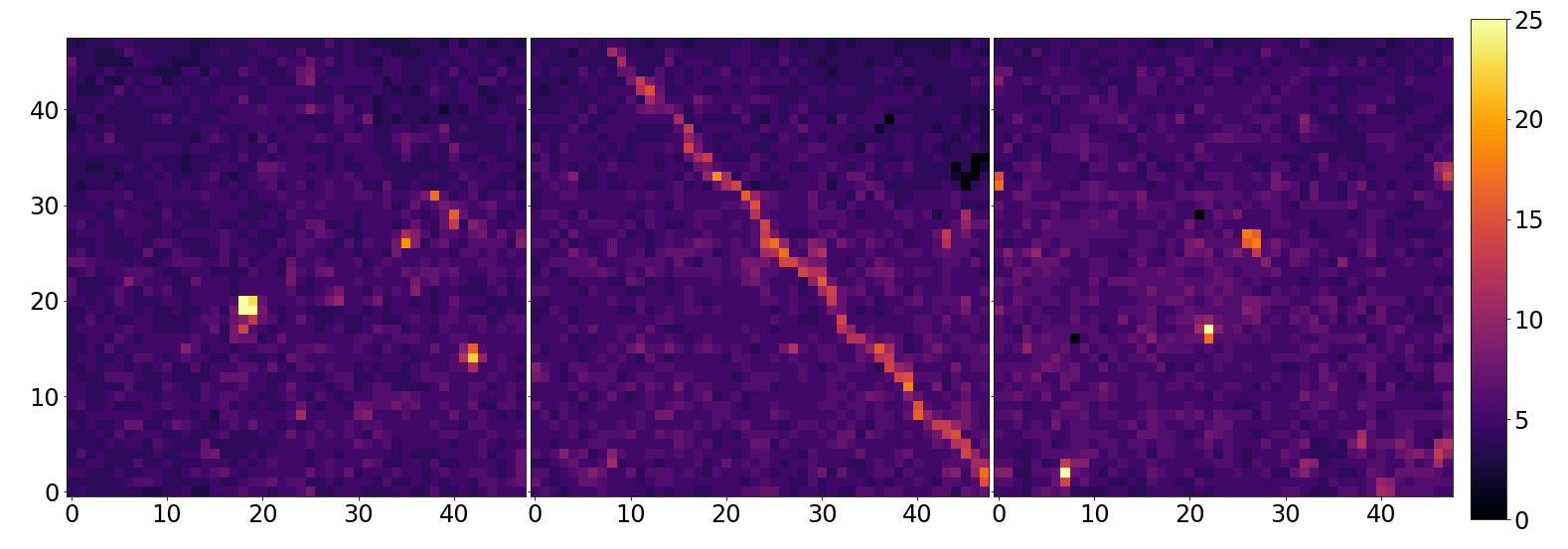}
\caption{Example integrated laser track recorded during the field tests on 31/08/2022. The color bar shows the photoelectron count on the focal surface. The laser energy was set to 1~mJ, and the laser at a distance of 24~km was pointed at 22\textdegree ~towards the detector while the telescope was pointed upwards. The bright pixels are stars inside the field of view.}
\label{fig:FT_filed_laser}
\end{figure}
\textbf{The Field tests} were conducted in July/August 2022 at the Telescope Array site in Delta, UT, USA, before the flight to characterize the fully assembled instrument. During the 12-day campaign, thousands of laser shots (see Fig. \ref{fig:FT_filed_laser} for an example track) and LED events at various monitoring conditions were recorded, providing an end-to-end photometric calibration (confirming the expected overall efficiency of the detector of 19\%), a field of view measurement and an energy threshold estimation. An in-depth discussion of the field test and the FT calibration is presented in \cite{ViktoriaICRC}. The obtained data set was also used to enhance the training set for the neural network used for event prioritization.

\begin{figure}
\centering
\includegraphics[width=.75\textwidth]{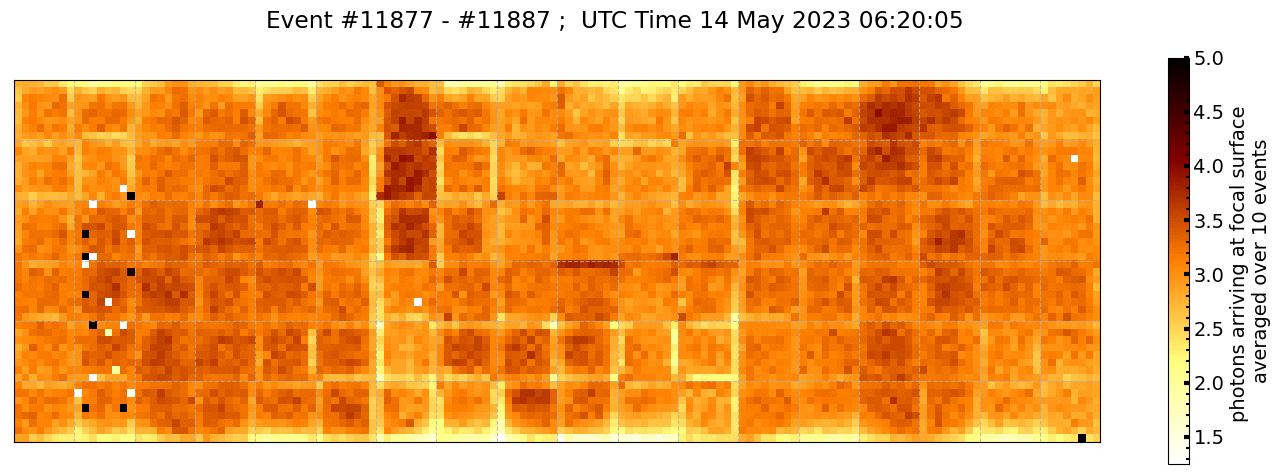}
\caption{An example of data recorded during the flight. The color scale shows the photon arriving at the aperture after applying the calibration to the raw count data. A total of 10 events (each consisting of 128 frames) were averaged. The dark parts are clouds moving through the instrument's field of view, which was confirmed by the corresponding IR picture. the black and white pixels are known malfunction pixel due to electronics issues.}
\label{fig:CloudFT.png}
\end{figure}
\textbf{The flight} lasted only for two data-taking periods, where the first one was mainly used for instrument commissioning after launch. However, we recorded and downloaded over 120,000 events over these two nights (e.g. Fig. \ref{fig:CloudFT.png}). 

We are currently performing various searches for the signal of an EAS induced by a UHECR, ranging from hand searches to employing different machine-learning approaches. To date, we have not found any UHECR candidate, which is consistent with the previously discussed simulation, which predicted one event in approximately 10 hours, assuming clear sky conditions, while we had less then 11 hours of data taking during the flight including almost 7 hours during the commissioning phase. More details of the flight timeline as well as preliminary analysis results, are presented in \cite{GeorgeICRC}.
The short flight still provides a lot of interesting events that will help to improve any future detector even further and to understand potential backgrounds better, as this is the first multi-PDM detector built. It will also be useful to develop a new trigger mechanism to reduce background triggers while retaining an energy threshold as low as possible to guarantee the detection of an EAS in the next flight; one such trigger could be a smart trigger based on a machine learning approach which is under investigation right now.
\section{The Cherenkov Telescope (CT)}
\label{sec:CT}

\textbf{The instrument} is also a 1m diameter modified Schmidt telescope with a bifocal alignment of the 4 mirror segments meaning light is focused in two distinct spots on the camera instead of one, helping to distinguish between direct cosmic ray hits (only one spot) and light from outside the telescope (2 spots), thereby reducing the background. The core of the Cherenkov telescope is a 512 Silicon PhotoMultiplier (SiPM: S14521-6050AN-04 from Hamamatsu) pixel camera with an integration time of 10~ns and a readout depth of 512 frames centered around the trigger, allowing to target very fast and bright signals, like the Cherenkov emission from air showers to achieve the goals outlined earlier. The FoV of the instrument is 6.4$^\circ$ in zenith and 12.8$^\circ$ in azimuth and can be pointed during the flight from horizontal to 10$^\circ$ below the Earth's limb depending on the targeted science. A more in-depth discussion of the instrument can be found in \cite{SPB2-CT, OscarICRC}.

\begin{figure}
\centering
\begin{minipage}{0.45\textwidth}
    \centering
    \includegraphics[width=.9\textwidth]{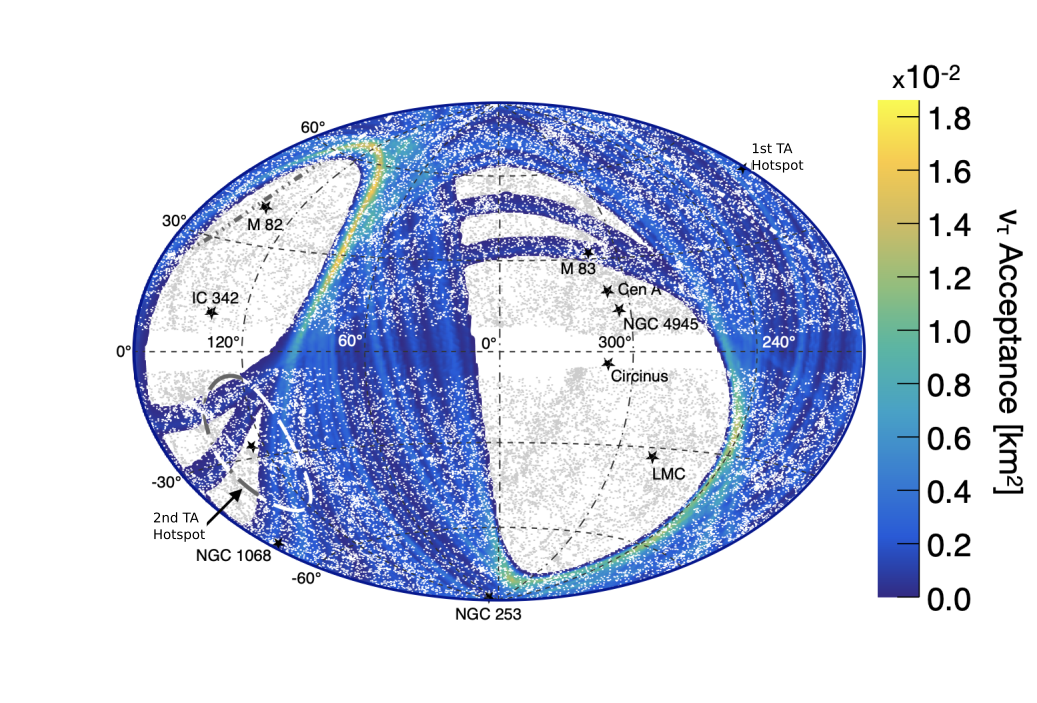}
    \caption{Expected average tau neutrino acceptance at $10^{8.5}$~GeV over 100 days, with a launch date of April 1, 2023, is shown in the sky plot, including the effect of the sun and the moon but no balloon motion.}
    \label{fig:SkyMapAcceptance}
\end{minipage}\hfill
\begin{minipage}{0.45\textwidth}
    \centering
    \includegraphics[width=1.\textwidth]{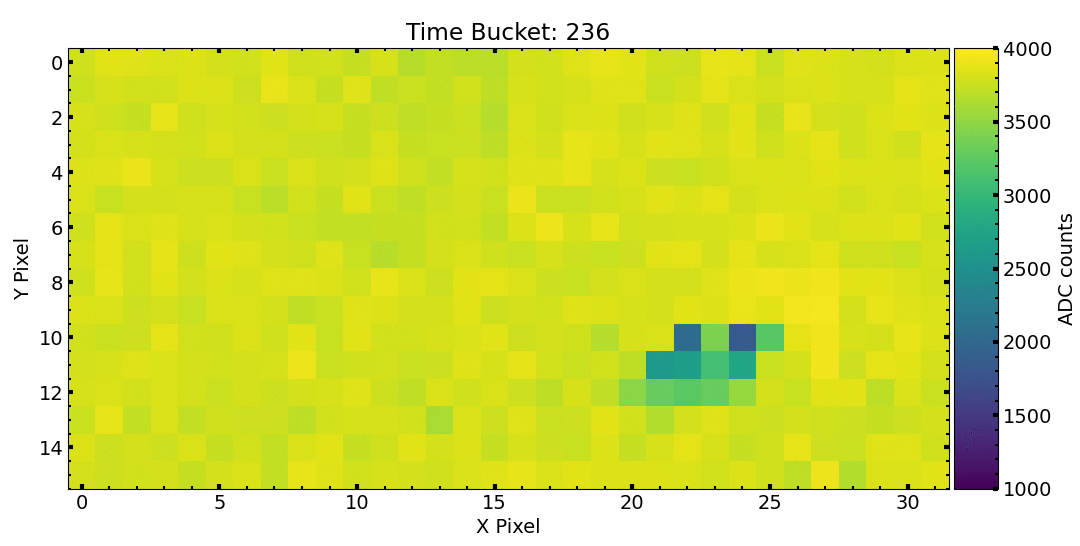}
    \caption{One frame of a downward going cosmic ray as observed by the CT of EUSO-SPB2 during field tests. The bifocal focus is clearly visible.}
    \label{fig:FieldCR}
\end{minipage}
\end{figure}

\textbf{The expected performance} of the Cherenkov telescope was also estimated before the flight with a wide range of simulations to cover the different scientific goals. For the above the limb cosmic rays, a full Monte Carlo analysis was performed where the \texttt{EASCherSim}\footnote{https://c4341.gitlab.io/easchersim/} code was used to generate the optical Cherenkov signal of showers, taking into account geometric and atmospheric effects \cite{Cummings_2021} and using trigger threshold obtained during field testing (see next paragraph). The most recent estimates also take the effect of the Earth's magnetic field into account and was adjusted for the unexpected lower balloon altitude of 18~km (details in \cite{AustinICRC}). The resulting expected rate is around 4 per hour accumulated over all energies with an energy threshold at around 10 PeV. The simulations also showed that the majority of PeV-scale events are observed a few degrees above the limb, but more energetic primaries have observable trajectories closer to the Earth.\\
Besides measuring backgrounds, the main scientific goal for below the limb observations was the so-called target of opportunity measurement, as it could be shown in simulation that EUSO-SPB2 has sufficient sensitivity to neutrinos from such events if they happen in our cosmic neighborhood \cite{Reno:2021xos}. An acceptance map is shown in Fig.~\ref{fig:SkyMapAcceptance}. A full machinery was developed to follow up on such alerts and to prioritize them to maximize the chances of neutrino observations (for details, see \cite{TobiasICRC, HannahICRC, JohnathanICRC}.

\textbf{The field test} was an essential part also in the flight preparation of the CT. They were conducted over 9 days in Jan/Feb 2022 at the Telescope Array site in Delta, UT, USA. During this time, the instrument recorded thousands of artificial light sources, i.e. laser tracks and LED flashes at various atmospheric and environmental conditions. In addition, the team recorded the Cherenkov emission from downward-going cosmic rays. An example of such an event is shown in Fig. \ref{fig:FieldCR}. While a full analysis is still required, it is reasonable to state that the measured rate of cosmic ray candidates is a good estimate of the rate of downward-going showers, therefore providing one measure of their validity.

\begin{figure}
\centering
\includegraphics[width=1.\textwidth]{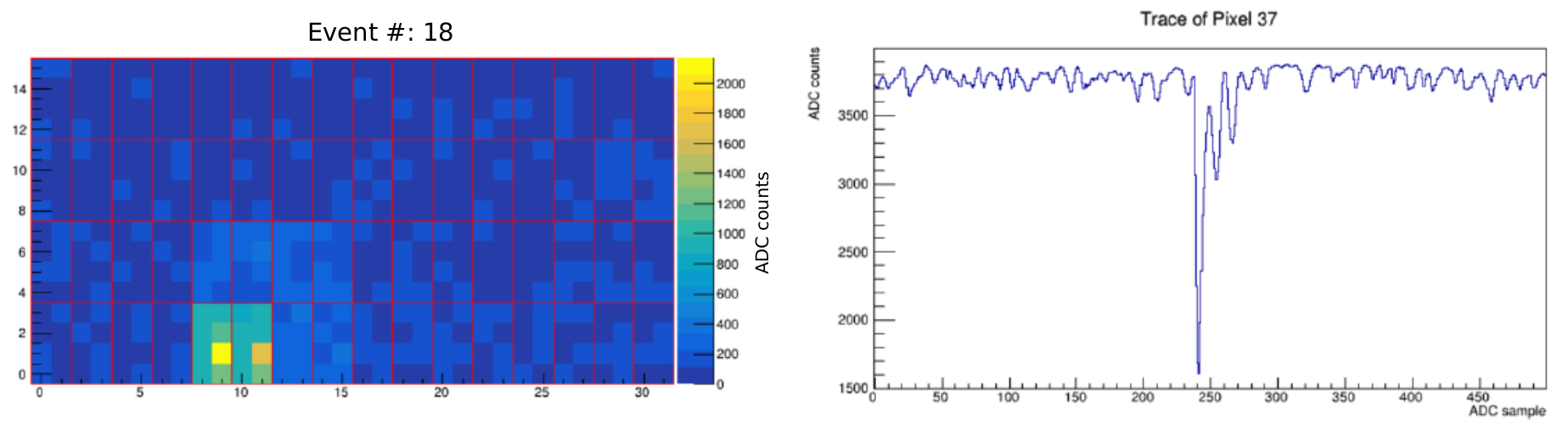}
\caption{An example of a cosmic ray event above the limb observed on May 14th 2023. The left panel shows a single frame of the entire camera, where color represents the ADC counts. The right panel displays a time trace of the pixel with the highest signal, indicating the intensity of the Cherenkov radiation from the cosmic ray event.}
\label{fig:FlightCR}
\end{figure}

\textbf{The Flight data} comprises camera performance data, night sky background images, and bifocal events collected during two nights of observation. The first night was mainly used for commissioning the instrument, which performed as expected at suborbital space for the first time. This demonstrates its potential for future space missions such as POEMMA. A detailed discussion of the instrument's flight performance and a preliminary summary of the data collection are given in \cite{ElizaICRC}. We manually scanned some of the more than 31,000 events triggered by the bifocal trigger to identify their possible origins. These bifocal events could be examples of extensive air showers (EASs) induced by cosmic rays or neutrinos, direct hits, or accidental triggers caused by background fluctuations. For about 45 minutes, we pointed the telescope above the Earth's limb to search for cosmic rays and found several candidates, one of which is shown in Fig. \ref{fig:FlightCR}. The left panel shows one frame of the entire camera, clearly demonstrating the bifocal alignment, while the right panel shows the signal trace of one of the pixels, emphasizing the signal strength of such an event.

For the rest of the time, we pointed the telescope below the limb to search for Earth-skimming VHE-neutrinos, but we did not find any candidates. This is consistent with the limited observation time and the lack of astrophysical alerts to follow up on. We will analyze the data more thoroughly in the future to characterize the night sky background and classify all triggered events. We also noticed some events that do not show bifocal topologies, which may be explained by direct cosmic ray hits.
\section{Summary \& Outlook}
\label{sec:summary}

EUSO-SPB2 was a mission of opportunity on a NASA super pressure balloon that launched successfully on May 13th from Wanaka, NZ. However, the balloon developed a hole in the envelope and was terminated over the Pacific Ocean after only about 37 hours of flight. Despite this very short flight, all instruments (FT, CT, and IR) worked as expected based on extensive simulations and rigorous laboratory and field tests. We downloaded around 56GB of data, which included more than 120,000 FT events and more than 32,000 bifocal CT triggers.

We are still analyzing this data set, but so far, we have not found any cosmic ray candidate in the FT triggers, which is consistent with the expected rate of less than one for the data taking time of less than 10 hours.

The CT data set contains triggers from below the limb, which enable us to study the backgrounds of potential neutrino observation for future missions and to demonstrate the detector's stability over the flight, which was one of the main mission goals. The observation above the limb for approximately 45 minutes resulted in the detection of several Cherenkov signals from upward-going EAS caused by cosmic rays, proving the feasibility of not only the developed trigger scheme but also the detection technique itself for the first time (another main mission objective). The short observation time will not allow a detailed statistical analysis of these events. We could not follow up on any target of opportunity during this flight duration, but the alert mechanism was working.

Encouraged by the instruments' performance, we started planning a new pathfinder balloon mission called POEMMA-Balloon with Radio. This mission will be designed with all the experience of EUSO-SPB2 in mind but with some significant changes and upgrades. The main change is to combine both telescopes into one with a hybrid focal surface similar to the original POEMMA design and to increase the light collection area by a factor of 1.2 to 1.5. In addition to the optical telescope, the PBR design will include a radio detector for the detection of upward-going EAS from cosmic rays and neutrinos to explore the synergy between the optical Cherenkov measurement and the radio detection of such events. The project is in the early design phase but has a target launch date of Spring 2026.
\section*{Acknowledgment}
\noindent
\small{
The authors would like to acknowledge the support by NASA award 11-APRA-0058, 16-APROBES16-0023, 17-APRA17-0066, NNX17AJ82G, NNX13AH54G, 80NSSC18K0246, 80NSSC18K0473, 80NSSC19K0626, 80NSSC18K0464, 80NSSC22K1488, 80NSSC19K0627 and 80NSSC22K0426, by the French space agency CNES, and by National Science Centre in Poland grant n. 2017/27/B/ST9/02162. This research used resources of the National Energy Research Scientific Computing Center (NERSC), a U.S. Department of Energy Office of Science User Facility operated under Contract No. DE-AC02-05CH11231. We acknowledge the ASI-INFN agreement n. 2021-8-HH.0 and its amendments. We acknowledge the NASA Balloon Program Office and the Columbia Scientific Balloon Facility and staff for extensive support. We also acknowledge the invaluable contributions of the administrative and technical staffs at our home institutions.}

\bibliography{icrc2023_bib}

%
\newpage
{\Large\bf Full Authors list: The JEM-EUSO Collaboration\\}

\begin{sloppypar}
{\small \noindent
S.~Abe$^{ff}$, 
J.H.~Adams Jr.$^{ld}$, 
D.~Allard$^{cb}$,
P.~Alldredge$^{ld}$,
R.~Aloisio$^{ep}$,
L.~Anchordoqui$^{le}$,
A.~Anzalone$^{ed,eh}$, 
E.~Arnone$^{ek,el}$,
M.~Bagheri$^{lh}$,
B.~Baret$^{cb}$
% ...
}
\end{sloppypar}
\vspace*{.3cm}

{ \footnotesize
\noindent
% ...
$^{cb}$ Universit\'e de Paris, CNRS, AstroParticule et Cosmologie, F-75013 Paris, France\\
% ...
$^{ed}$ Istituto Nazionale di Fisica Nucleare - Sezione di Catania, Italy\\
% ...
$^{eh}$ INAF - Istituto di Astrofisica Spaziale e Fisica Cosmica di Palermo, Italy\\
% ...
$^{ek}$ Istituto Nazionale di Fisica Nucleare - Sezione di Torino, Italy\\
$^{el}$ Dipartimento di Fisica, Universit\`a di Torino, Italy\\
% ...
$^{ep}$ Gran Sasso Science Institute, L'Aquila, Italy\\
% ...
$^{ff}$ Nihon University Chiyoda, Tokyo, Japan\\ 
% ...
$^{ld}$ University of Alabama in Huntsville, Huntsville, AL, USA\\
$^{le}$ Lehman College, City University of New York (CUNY), NY, USA\\
% ...
$^{lh}$ Georgia Institute of Technology, USA\\
% ...
}

\end{document}